# Nanophotonic materials for space applications

Ognjen Ilic, Department of Mechanical Engineering, University of Minnesota, Minneapolis, MN, USA; ilic@umn.edu

Space exemplifies the ultimate test-bed environment for any materials technology. The harsh conditions of space, with extreme temperature changes, lack of gravity and atmosphere, intense solar and cosmic radiation, and mechanical stresses of launch and deployment, represent a multifaceted set of challenges. The materials we engineer must not only meet these challenges, but they need to do so while keeping overall mass to a minimum and guaranteeing performance over long periods of time with no opportunity for repair. Nanophotonic materials—materials that embody structural variations on a scale comparable to the wavelength of light—offer opportunities for addressing some of these difficulties. Here, we examine how advances in nanophotonics and nanofabrication are enabling ultrathin and lightweight structures with unparalleled ability to shape light-matter interactions over a broad electromagnetic spectrum. From solar panels that can be fabricated in space to applications of light for propulsion, the next generation of lightweight and multifunctional photonic materials stands to both impact existing technologies and pave the way for new space technologies.

**Introduction**

The convergence of technology advances and public and private interests is opening access to space to an unprecedented degree. Launch costs, among the primary constraints on space missions, have been substantially reduced through reusable rockets, vertical integration of production, and market competition.[1] The concept of what a spacecraft could look like is rapidly evolving, with the proliferation of miniature satellites—including CubeSats assembled from 10 cm x 10 cm x 10cm cube units (**Figure 1**c)—that are cheaper to manufacture, launch, and deploy in large numbers.[2–5] Built with the same off-the-shelf hardware that goes into our cell phones, these small space objects leverage the technological leaps in nano- and micro-processing, integrated circuitry, and economies of scale across nanoelectronics industries.

However, the nanotechnology revolution that made these concepts possible has not been limited just to our ability to control the flow of electrons on the smallest of scales. It has led to materials breakthroughs for lighter, more resilient, and multifunctional components and materials systems with tailored properties especially suitable for the space environment.[6–8] It has also enabled us to shape how light interacts with matter across the electromagnetic spectrum and has paved the way for radically different optical concepts and devices (Figure 1a, b). This article discusses several ways in which advances in understanding light-matter interactions at the nanoscale could impact some of the existing space technologies, as well as lead to new, potentially transformative applications.

Nanophotonic materials and metamaterials can exhibit highly unusual optical responses not found in natural materials. Properties such as negative refraction, anomalous dispersion, and the existence of photonic bandgaps arise from subwavelength features and could be exploited in a number of applications, including for beam steering, reconfigurable optics, cloaking, imaging, photonic signal processing, and many others.[9] Depending on the application, photonic designs at the nanoscale have evolved to encompass a wide variety of motifs, from multilayer stacks and two- (2D) and three-dimensional (3D) photonic crystals, to the more recent planar aperiodic meta-atoms and metasurfaces.[10,11] In conjunction with these developments, advances in materials synthesis and fabrication in the last decade—including single- and few-atomic-layer materials[12]—have enabled unprecedented control of material arrangements and stacking down to the nanometer level. This has opened rich opportunities for combining material properties with structural patterns, and emerging sophisticated photonic design methods can navigate the vast landscape of possible combinations.[13] Importantly, such rich functionality can be realized in ultrathin and lightweight structures, enabling mimicking of bulk optical elements in a substantially smaller form-

factor—particularly desirable for applications in space where there is a substantial cost incentive to keep the payload weight as low as possible.

Of the many applications of light in space, this article narrows its focus to the challenges and opportunities of controlling radiative flows of energy, that is, energy associated with sunlight, heat, and propulsion. We first review photovoltaic harvesting of solar energy in space and discuss opportunities for novel materials for solar panels that are efficient, flexible, and foldable, and could even be fabricated in space. Radiative heat is another dominant form of energy. In the absence of conduction and convection in the vacuum of space, heat needs to be dissipated radiatively, and this article discusses ways in which nanophotonic control of thermal emission could assist with thermal management, as well as enable efficient conversion of heat into usable electricity. The energy and the momentum of light, whether from the sun or a laser, can also propel objects in space. Depending on the size and shape of the object, this effect can range from a gentle nudge to substantial acceleration. We examine how photonic metamaterials and metasurfaces can impact light-based propulsion for space travel within and beyond our solar system. Finally, we reflect on the challenges of materials testing and concept demonstrations at scales suitable for practical applications.

**Harnessing sunlight in space**

The history of using the energy of sunlight in space is almost as old as human-made satellites themselves, starting with the launch of *Vanguard I* in the 1950s, the fourth satellite to be placed in Earth orbit and the first fitted with solar cells. The quality, efficiency, and reliability of photovoltaic elements used in space has since dramatically improved,[14] following similar trends of material development for sunlight harvesting on Earth. Yet, the notion of what constitutes a desirable photovoltaic technology differs radically between terrestrial and space environments. On Earth, techno-economic considerations tend to favor inexpensive solar cells that are moderately efficient but cost-effective to produce, deploy, and maintain at scale. In contrast, a solar panel in space needs to be lightweight and highly efficient, but also resistant to harsh environmental conditions and reliable for long-term operation with minimal-to-no opportunity for repair. These considerations have led to the adoption of an alternative figure of merit of specific power (W/kg), rather than specific cost (W/$), for evaluating photovoltaic technologies in space. The most efficient solar panels presently used in space are multijunction gallium arsenide cells,[15] with conversion efficiencies around 30%. Ongoing research and design improvements could boost efficiencies to >37% and beyond.[16,17,18]

In terms of specific power, novel hybrid and organic photovoltaic cells under development can significantly outperform existing photovoltaic technologies. In particular, perovskite solar cells—which draw their name from the perovskite-structured light-harvesting layer—have dramatically improved in performance during the last decade, with tandem configurations reaching power conversion efficiencies of 26.7%.[19] A highly absorbing active layer enables sunlight to be captured within an ultrathin, submicron layer, ultimately leading to high specific powers exceeding 20 W/kg.[20] It is not just the potential for record-setting specific powers that makes perovskite solar cells promising candidates for applications in space. They can be made flexible and foldable for easier transport, potentially reducing costs and challenges associated with launch and deployment. The fabrication process can be inexpensive in relative terms and, crucially, perovskite panels have the potential to be fabricated in space.[21] The ability to transport the material into space in condensed form (e.g., powder or a liquid) and then print it onto panels could be revolutionary, not just for assembly but also for repair. There are exciting opportunities for nanophotonic engineering to further improve the performance of perovskite cells, including enhanced light trapping by diffraction,[22] and light management for the suppression of reflection losses.[23,24]

Despite the disruptive potential of these novel ultra-thin and ultra-lightweight photovoltaic materials, there has only been limited testing of how they might respond to the harsh conditions in space. For example, high mechanical stresses, extreme temperature cycles, and bombardment by high energy particles could all impact the efficiency of light absorption and charge transport.[25] Comprehensive additional testing of electro-optical properties is needed, though there are some promising preliminary results. In an attempt to mimic cosmic particle radiation, recent efforts have shown that perovskites could be stable in high-fluence proton beams and, in fact, exhibit radiation tolerance superior to that of silicon or gallium-arsenide cells.[26,27] Besides perovskites, nanowire solar cells represent another candidate technology wherein photonic engineering of the nanowire morphology enables effective trapping of sunlight in a thin absorber layer.[28] Early-stage efforts have shown promising potential for radiation tolerance in III–V nanowire solar cells and have provided design guidelines for further improvements in specific power for space applications.[29]

But power generated from sunlight does not have to be consumed in space; it could be sent to Earth. The idea of harnessing the full intensity of the sun in space, unaffected by day-night transitions and atmospheric losses, and then wirelessly transmitting it to the ground is more than half a century old.[30] Following early feasibility studies by the US Department of Energy and NASA in the 1970s, a number of variations of this concept have been proposed. Generally, the solar energy collected with high-efficiency photovoltaic panels would be wirelessly beamed back to Earth, either at laser or microwave frequencies. The advantage of using comparatively shorter laser wavelengths for beaming power is that transmitting and receiving apertures could be smaller (and thus cheaper). On the other hand, microwaves are less susceptible to atmospheric effects, and microwave power transmission has been extensively studied for applications in space.[31] All space solar concepts share a similar set of challenges: The components need to be lightweight but durable, compact but efficient, and cost-effective to deploy and operate. Moreover, for this method of power generation to be competitive, both modules (in space and on Earth) and their aperture sizes need to be large (~km in size). The combination of these extraordinary technical and, more importantly, economic challenges has constrained the development of this concept.

Advances in optical and nanostructured materials are providing us with fresh insights into addressing some of the issues of space solar harvesting. One example is a technology under development at the California Institute of Technology.[32] This approach envisions ultra-lightweight photovoltaic concentrator panels (**Figure 2**). The design is highly modular, consisting of an assembly of independent "tiles" capable of performing all functionality associated with solar power harvesting: the tiles collect sunlight and convert it into electricity, transform energy into longer wavelength radiation, and finally beam it away via a phased array of antennas.[33,34] A key optical design decision is the proposed thin-film solar harvesting panels with parabolic reflectors and edge-lined photovoltaic cells. It reduces the footprint of photovoltaic components, which minimizes mass as well as the area for radiation shielding. The parabolic reflector acts as a concentrator of sunlight (concentration factor ~10–20x), provides mechanical support and electrical wiring, and helps spread heat. The last point touches on the broader issue of thermal management in space, an area where novel optical materials and metamaterials can have a dramatic impact. Illuminated by concentrated sunlight, embedded solar cells heat up but are unable to release that heat by convection, as solar cells do on Earth.

Effective thermal management is crucial for ensuring that the photovoltaic cells can operate at optimal efficiencies, and nanostructured materials can play a vital cooling role by dissipating heat by emitting thermal radiation. Typical high thermal emissivity coatings used in space, such as paints with high $TiO_2$ or $Al_2O_3$ content, are diffuse reflectors for solar wavelengths and thus less suited for the mirror concentrator surface. In contrast, thin-film nanophotonic structures can be designed to be both specularly reflective (for sunlight) and thermally emissive, as recent examples based on ultra-lightweight layers of colorless polyimide 1 (CP1) and indium-tin-oxide have shown.[35,37] We examine the challenges and opportunities of using nanophotonic materials and metamaterials for heat dissipation in a separate section on radiative thermal management. Besides optical materials and components, technological improvements across the board—in power electronics, mechanics of assembly/deployment—are being investigated by Caltech, Northrop Grumman, US Naval Research Laboratory, the Japanese Space Agency (JAXA), and various NASA-funded projects.[31] Such materials advances, together with significant trends of reducing the cost of access to space, make the potential of some form of space solar power development more favorable than ever before.

**Electricity from heat**

Further away from the sun, solar panels become less effective at generating electricity. Radioactive isotopes are alternative sources of energy that can be carried on-board for missions that require an independent means of energy generation. In particular, plutonium-238 has high power density (~0.57 W/g) and long half-life (~87 years) which make it suitable for long-term energy delivery. This longevity has been demonstrated in numerous space missions, most famously in the Voyager program (*Voyager I* and *Voyager II*), where the power systems remain operational more than four decades since deployment. In these "space batteries," the heat released during radioactive decay is converted into electricity with a thermoelectric converter. The most widely used converters are radioisotope thermoelectric generators (RTGs) that operate on the principle of the Seebeck effect, whereby a temperature difference across a material induces a voltage (**Figure 3**). However, the efficiency of an RTG—the ratio of the generated electrical power per unit of thermal power of the fuel source—is typically limited to 4–6%, depending on the material's thermal/electrical conductivity, intrinsic Seebeck coefficient, and the operating temperature.[37,38] For further improvements in the conversion efficiency, better performing materials need to be developed—a research area of very active interest in the thermoelectric community, for both terrestrial and space applications.[39]

Radioisotope thermophotovoltaic (R-TPV) systems are an alternative heat-to-electricity harvesting technology that converts the energy of incandescent photons into electrical power, with the potential for significantly higher conversion efficiencies than RTGs. The principle of operation (Figure 3a) mimics that of typical solar cells, where instead of the sun and the solar cell, a high-temperature thermal radiation source is coupled with a low-bandgap photovoltaic cell. The energy from the radioactive decay is used to heat a thermal emitter, and the radiated photons are subsequently converted into electricity inside the photovoltaic element. A key difference between solar-photovoltaics and thermo-photovoltaics is the thermal radiator temperature and the corresponding choice of the photovoltaic semiconductor material. In contrast to widely used silicon solar cells, typical temperature ranges for thermal emitters (~1000–1500°C) limit this technology to semiconducting materials with a lower bandgap, such as GaSb, InGaAs, and InGaAsSb.[40]

The promise of thermophotovoltaic energy conversion lies in its potential for high specific power and high conversion efficiencies when the spectrum of thermal emission is tuned to the semiconductor band-edge absorption. For thermal emission from a black/gray body, only the photons of energy greater than the semiconductor bandgap can be converted, as photons with lower energy are unable to generate electron-hole pairs. Moreover, for photons that can be absorbed, the energy in excess of the bandgap energy is usually lost to thermalization. A way to overcome this inherent inefficiency is to tailor the spectrum of high-temperature thermal emission with nanophotonic materials (Figure 3). In particular, structures with features smaller than the wavelengths associated with thermal radiation offer broad opportunities for realizing emission tuned to the photovoltaic bandgap.

In the context of tailoring the spectrum of high-temperature radiation, selective thermal emitters have assumed a variety of photonic profiles,[40–42] including 1D layered structures,[43–45] 2D photonic micro-cavities,[46–48] as well as 3D woodpile and self-assembled structures.[49,50] Photonic crystal slabs with periodic holes etched into the surface of refractory metals (such as tungsten and tantalum) are a particularly promising material platform, owing to high-temperature stability and low emission in the far-infrared.[51] Their thermo-optical properties can be further enhanced with superlattice patterning for even greater design flexibility.[52] Reuse of unabsorbed sub-bandgap photons could be aided by highly reflective rear mirrors for a significant boost in performance.[53] Under idealized conditions, theoretical limits on the thermophotovoltaic conversion efficiency are very high (>80%), more than double the Shockley–Queisser limit of solar energy conversion.[54]

In practice, the broadband nature of thermal radiation limits the extent to which spectrally selective emission is possible, leading to parasitic and sub-band emission. Similarly, the intrinsic efficiency and quality of specialty low bandgap TPV cells has been relatively low compared to well-developed silicon solar cells.

Nevertheless, the potential for superior performance continues to drive theoretical and experimental efforts into material (e.g., better metamaterial emitters, higher quality multijunction TPV cells, better rear mirrors) and system-level design improvements. For applications in space, TPV offers potential improvements not just in efficiency but also specific power (W/kg)[55,56] and continues to motivate nanophotonic material development.[57]

**Radiative thermal management**

The engineering of thermal emission in space is relevant beyond the conversion of heat into usable electrical energy. In the absence of conduction and convection, radiative emission is the dominant mechanism for thermalization and the exchange of energy with the space environment. Here, the same material design principles for efficient conversion of heat into electricity also apply to tailoring thermal emission for radiative thermal management of objects and structures in space.

Radically different from conditions on Earth, the space environment presents a unique challenge for thermal management, where a structure can be exposed to the two extreme sources of thermal radiation: the hot sun at close to 6,000 degrees Kelvin and the cold space at only 3 degrees Kelvin. A typical approach to shield from the sun while dissipating internally generated heat is to use optical solar reflectors mounted on the exposed radiator panel surfaces. Solar reflectors are multi-band optical devices: they are engineered to have high reflectivity and low absorption at shorter wavelengths at the peak of the solar spectrum and high emissivity (i.e., low reflectivity) in the long-wave thermal infrared portion of the electromagnetic spectrum. Typical examples of solar reflectors are quartz tiles, where a top layer of quartz is mounted over a reflecting metallic layer. The two layers operate in tandem: the quartz layer efficiently emits thermal infrared radiation while transmitting solar light to be reflected from the underlying metallic surface. Despite their good thermo-optical properties, quartz tiles can be challenging to apply to flexible radiator panels and they also add extra weight.

Metamaterial and metasurface photonic engineering could provide a new path toward overcoming such challenges, with structures that can be made lightweight without sacrificing optical performance. One example is the class of transparent conducting oxide metasurfaces. Such photonic metasurfaces incorporate transparent material films in which the dopant density can be controlled to tune the material response across a broad spectral range associated with thermal emission. In a recent work, a combination of an aluminum-doped zinc oxide metasurface and a metallic back-reflector was shown to simultaneously exhibit the desired dual-band behavior of low solar absorptance (<0.2) and high thermal-infrared emissivity (>0.7).[58] Here, the principle of operation is based on a Salisbury screen configuration, where the metasurface layer is placed a quarter-wavelength away from the reflecting plane to maximize absorption. Nanoscale dimensions of the metasurface pattern, its thickness, and the properties of the spacer layer are carefully tuned to achieve optimal thermo-photonic properties. These proof-of-concept results can be further improved by suppressing parasitic ultraviolet absorption of sunlight, and one approach is to incorporate a thin-film multilayer ultraviolet-reflector.[58] Typically, the requirement for additional components can negatively impact mechanical as well as optical properties of the device. In this case, however, the multilayer reflector operates at short, ultraviolet wavelengths and can be made thin enough (<1 μm) to avoid adverse effects. From a broader perspective, this point illustrates the intrinsic strategy associated with tailoring thermal emission. The broadband spectrum of thermal radiation often means that best performing photonic designs are not monolithic in nature, but rather they are a combination of appropriate materials and nanoscale patterns.

Nanophotonic metamaterials and metasurfaces could also enable novel approaches for adaptive thermal management. During the course of a day, structures in space may experience intense differences and changes in heating rates depending on their orientation relative to the sun. As mentioned, these temperature swings can lead to material degradation and impact longevity, and need to be accounted for in material and component design. However, smart materials can respond to temperature changes to self-regulate the net balance between the emitted and the absorbed thermal radiation. For example, when the structure becomes very hot, a state of high thermal emissivity would be desirable to radiate the heat away;

as it cools down, low thermal emissivity is advantageous to prevent the temperature from becoming too low.

Phase change materials are especially appealing for thermal self-regulation without the need for external energy input or control. A potential candidate is vanadium dioxide ($VO_2$), a thermo-chromic material with a characteristic insulator-to-metal phase transition at the critical temperature $T_c$ (~70°C). For temperatures below $T_c$, the insulating state of $VO_2$ is characterized by higher resistivity and a smaller imaginary part of the refractive index; for temperatures above $T_c$, the metallic phase exhibits a lower resistance and higher absorption coefficient. By themselves, the two phases of $VO_2$ are not especially useful. However, by integrating $VO_2$ within an engineered nanophotonic environment, the optical contrast between the two phases can be tailored and amplified, and used to stabilize the temperature.

Applying this reasoning toward adaptive thermal management, recent works have proposed the use of phase-change vanadium dioxide ($VO_2$) integrated into a metamaterial tile,[59] thin-film,[60] or a cone[61] configuration. The purpose of the metamaterial integration is to enhance emissivity in the high-temperature state of $VO_2$ and, correspondingly, suppress emissivity in the low-temperature state. Proof of concept results are encouraging; for example, a change in emissivity of 0.48 was recently measured,[59] and a large emissivity difference of 0.8 between low and high states is predicted.[61] Further improvements can be obtained with the use of low-emissivity dielectric spacer materials and more sophisticated photonic design. In addition, the material properties of $VO_2$ films can be manipulated (e.g., via defect-engineering[62]) to shift the critical phase-change temperature and the operable temperature range. It is less likely that such passive thermal management approaches could rival the capability of active systems that employ temperature sensors and electronic controls, yet their advantage lies in their simplicity. They have no moving parts, require no supply of power, and are ultra-lightweight: the functional part of such metamaterials and metasurfaces can be made remarkably thin (~1–10 µm).

**Photonic materials for propulsion**

Lossless propagation of light over long distances opens opportunities for non-traditional means of space propulsion. The idea of harnessing the energy and momentum of light for accelerating objects in space is not new. Curiously, propulsion was among the earliest proposed applications for laser light[63–65] in the immediate aftermath of the invention of the solid-state laser in 1960.[66] The concepts of light-driven (and light-assisted) propulsion have evolved over time with several demonstrations of using laser, microwave, and solar energy for acceleration.[67–70] Even more so than for other space technologies, the effectiveness of photon propulsion critically depends on the ability to use lightweight, resistant, and multifunctional materials. These stringent material requirements lead to spacecraft concepts that are fundamentally different from the kind of spacecraft that are typically launched into space. Recent advances in understanding the physics of light-matter interactions at the nanoscale are pointing toward novel materials design principles for efficiently accelerating objects in space.

Depending on whether the propellant is carried onboard, there are two main mechanisms of light-based propulsion. For propellant-based systems, light is used to eject material (propellant) to provide thrust. While conventional rockets and spacecraft carry a propellant, light can provide an external source of energy for thrust generation, which eliminates the need for carrying and accelerating an onboard power supply. There have been several concept examples and demonstrations, such as producing thrust via light-induced ablation[71,72], microwave thermal heat-exchange,[68,73,74] emission of high-energy protons and ions,[75,76] and others.[70,77] In these systems, it is important to optimize photon-matter interactions to achieve efficient absorption and energy delivery. One novel approach that seeks to exploit resonant photonic phenomena is the concept of plasmonic propulsion,[78] where light focused onto deep-subwavelength nanostructures can induce plasmonic forces to accelerate and expel nanoparticles. The thrust generated in this manner can be amplified in a multi-layered, multi-stage configuration of nanostructures. Furthermore, by tailoring the nanoscale dimensions of each structure, a multi-resonant system can be designed to harness a broadband light source, such as sunlight. Such concepts could further benefit from related proposals of using nanostructured materials to tailor electromagnetic fields for the efficient laser-powered acceleration of charged particles.[79]

In the case of sunlight, it's not only the energy, but also the momentum of photons that can be harnessed for propulsion. When sunlight reflects from (or gets absorbed by) an object in space, it imparts a tiny amount of momentum that exerts pressure. Typically, solar radiation pressure is weak. Nevertheless, it is persistent, and large, sheet-like structures can exploit it for steady acceleration (**Figure 4**). Such solar sails[80] have been demonstrated on several recent occasions, including IKAROS,[69] NanoSail-D, LightSail (1 & 2),[81] and others,[82] with additional missions planned or under way.[83] The extraordinary length-to-thickness ratio in these structures—e.g., greater than $10^6$ for the IKAROS sail consisting of aluminum-coated polyimide sheet only 7.5 um thick and 20 m across—demands that all components, including sensors, photovoltaic cells, and attitude controllers, be in thin-film form. Here too, there are opportunities for nanophotonic materials to improve existing and to enable new capabilities. Novel photovoltaic technologies under development previously discussed could also find use in next-generation solar sails. For attitude control, engineered nanostructured layers that incorporate phase-change materials or 2D-material heterostructures[84] could offer superior reflectivity modulation relative to moderately performing liquid-crystal panels, such as those on the IKAROS sail. Similarly, the magnitude and the direction of solar radiation force can be tailored in a fundamentally novel manner. If the surface of the sail is patterned with arrays of nano-engineered structures—for example diffraction gratings or phase-gradient metasurfaces—then light could be reflected in any desired direction, beyond simple specular reflections, such as those from a polished surface.[85–92] By "steering" light, such sails could steer themselves, opening up exciting opportunities for trajectory control and maneuvering capability in space (**Figure 5**).

Laser radiation could overcome the limitations of using sunlight for propulsion. Lasers can provide orders-of-magnitude stronger light intensity and, unlike sunlight, are monochromatic—meaning a light sail would need to be reflective only for a single wavelength rather than a broad spectrum. Recently, there has been a renewed interest in the concept of laser-propelled ultrathin structures that can be accelerated to very high speeds. In particular, the Starshot Breakthrough Initiative envisions launching a spacecraft capable of reaching Proxima Centauri b—an exoplanet within the habitable zone of the star Proxima Centauri, about 4.2 light years away from Earth—a flight that would take approximately 20 years.[93–95] The spacecraft would consist of a thin-film lightsail propelled by a powerful array of ground-based lasers and a payload with the electronics and sensors to gather data and transmit it back to Earth. To traverse the distance to Proxima b in this time frame, the lightsail would need to attain relativistic speeds (e.g., ~60,000 km/s or 20% the speed of light), requiring gigawatt-power laser arrays over a kilometer in size.[95,96] These are extraordinary demands, yet the motivation for laser propulsion is fundamental. For a rocket, or any other object that accelerates its own fuel, no amount of propellant would be practical for such high-speed spaceflight since the maximum speed gain is only logarithmic with added fuel mass; for example, a spacecraft carrying $10^{10}$ times more fuel than payload could only experience ~20x velocity gain relative to the velocity of the exhaust (typically ~10s km/s). Simply put, there is no viable technology other than laser propulsion that could get us closer to the nearest stars in a timeframe that is compatible with the human lifespan.

The extreme engineering challenges of laser-propulsion offer a manifold of opportunities for nanophotonic materials. To maximize thrust, it is desirable that the reflectivity of the sail be as high as possible. Previous studies have argued for the use of dielectric thin-film multilayers[95,97] since these films can be made as reflective as metals but have significantly higher laser damage threshold. However, more sophisticated nanophotonic structures, such as photonic crystals, could perform even better than thin films while further reducing mass.[98] Besides being highly reflective, the sail material needs to minimize the absorption of laser light and maximize emission in the long-wave infrared, so that it could balance the temperature by radiative cooling. Toward that goal, multi-material heterostructures can combine properties from a combination of materials, for example, silicon for its high optical contrast and silica for its high emissivity via infrared phonon-polaritons,[99] to simultaneously facilitate efficient propulsion *and* radiative thermal management. In addition, to maintain its position and orientation during propulsion, it is necessary that a sail can stably ride its beam.[100,101] Here too, photonic meta-gratings and metasurfaces can offer a pathway for controlling the dynamics of flight (Figure 5a). For example, a flat, specularly reflective sail would be knocked off the beam axis by any small perturbation; however, a sail with surface-embedded nanoscale patterns could self-stabilize.[88–92,102] These meta-optical elements can be engineered to scatter light in a way that re-orients and pins the sail to the beam axis (Figure 5), an example of how nanophotonic materials can provide creative new solutions for opto-mechanical stability.

The excitement around the light-based propulsion concept is that it can be an intrinsically scalable technology. The vision of ultrahigh speed spaceflight necessitates high-power kilometer scale arrays, large lightsails, and the expansive supporting infrastructure that would accompany such an endeavor. However, a significantly scaled-down version of the concept could still find use in novel missions for space exploration of our own solar system (Figure 4b). Curiously, smaller, wafer-like sails propelled by a moderately powered (~100 kW), meters-sized laser array could be quickly accelerated to speeds characteristic of record-setting human-made spacecraft,[95,96] such as the *New Horizons* mission to Pluto and the Kuiper belt. Such scalable spaceflight technology could enable truly exciting opportunities for "hopping" around our own solar system, sending and guiding swarms of miniature space crafts to sense and observe our cosmic neighborhood in a radically new way. From searching for biological molecules in the plumes of Saturn's moon Enceladus[103] to intercepting interstellar "visitors" such as the mysterious *'Oumuamua* object that recently went past Earth,[104] the vast cosmic distances could be made dramatically shorter, and perhaps missions to deep space could become routine.

**Conclusion**

Breakthroughs in our understanding of light-matter interactions at the nanoscale are poised to dramatically shape new technologies in space, impacting energy harvesting, thermal management, and propulsion, as well as areas of communication and sensing[105–107] that are beyond the scope of this article. Photonic materials and devices discussed here are early-stage efforts and examples of proof-of-concept structures evaluated primarily in a laboratory environment. Significant further development and testing is needed toward their ultimate application in space missions. For many materials, it is not just optical and electromagnetic but also mechanical, chemical, thermal, and other properties that would need to be engineered in unison for the most ambitious space missions.[6,7]

In addition to challenges associated with dependable and long-term operation in the hostile space environment, considerations of cost-effective fabrication, integration, and assembly are particularly important. It is common for nano-optical structures to be assembled using fabrication methods that are primarily used for small-scale prototypes (e.g., atomic layer deposition, electron-beam lithography). Demonstrating the discussed nano-optical concepts over large areas is a key step toward applications, and there are significant ongoing developments aimed at advancing high-fidelity and cost-effective nanofabrication at scale, including nano-imprint and roll-to-roll lithography.[108,109] It is worth pointing out that considerations of cost and scalability also stand in the way of many commercial uses of nanophotonic materials here on Earth. Applications in space, where premiums are placed on high performance and low weight, could thus be near-term commercial drivers for materials and process development, a stepping stone toward widespread adoption of these breakthrough technologies on Earth and in space.

**Figures**

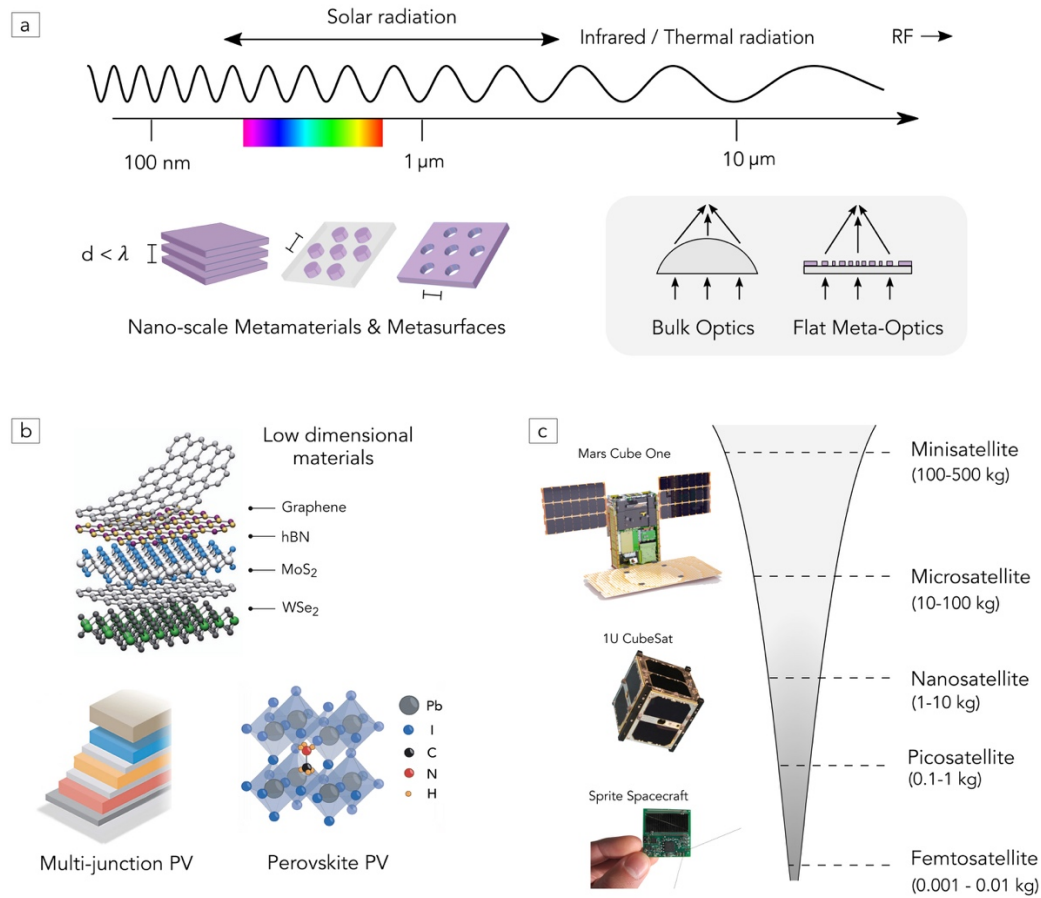

**Figure 1.** (a) Nanoscale photonic structures, such as metamaterials and metasurfaces, draw their properties from patterns of subwavelength elements. They enable diverse optical functionality over a broadband electromagnetic spectrum. (b) Examples of materials advances toward the control of light-matter interactions in extraordinarily thin and lightweight structures: heterostructures of two-dimensional materials with wide-ranging optoelectronic and optomechanical properties. Many-junction solar cells and emerging photovoltaic technologies for high efficiency sunlight harvesting in space. (Reproduced with permission from Reference 12.) (c) The ongoing trend towards reducing the form-factor of objects launched in space: small satellites (< 200 kg). Examples shown are Mars Cube One (MarCO-A and B) satellites, one-unit CubeSat, and wafer-scale spacecrafts. Ultrasmall spacecraft could operate as a collaborative swarm; these are inexpensive to build and deploy in large numbers. (Credit: NASA/JPL-Caltech, Z. Manchester).

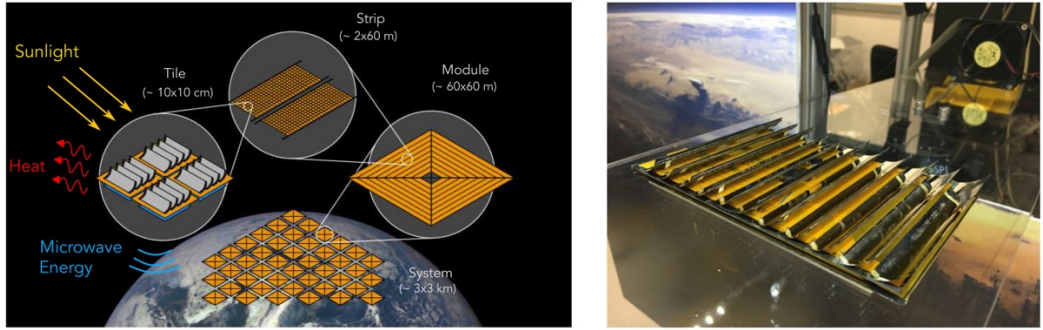

**Figure 2.** Harvesting solar energy in space. A concept of a space solar power station to provide dispatchable energy to a terrestrial receiver, unaffected by nighttime, seasonal variations or geographical location. The modular design consists of lightweight materials with multifunctional electromagnetic functionality: collect sunlight, radiate heat, and transmit radio waves. Right: prototype tile under development.[32-34] (Courtesy of Space Solar Power Project, Caltech)

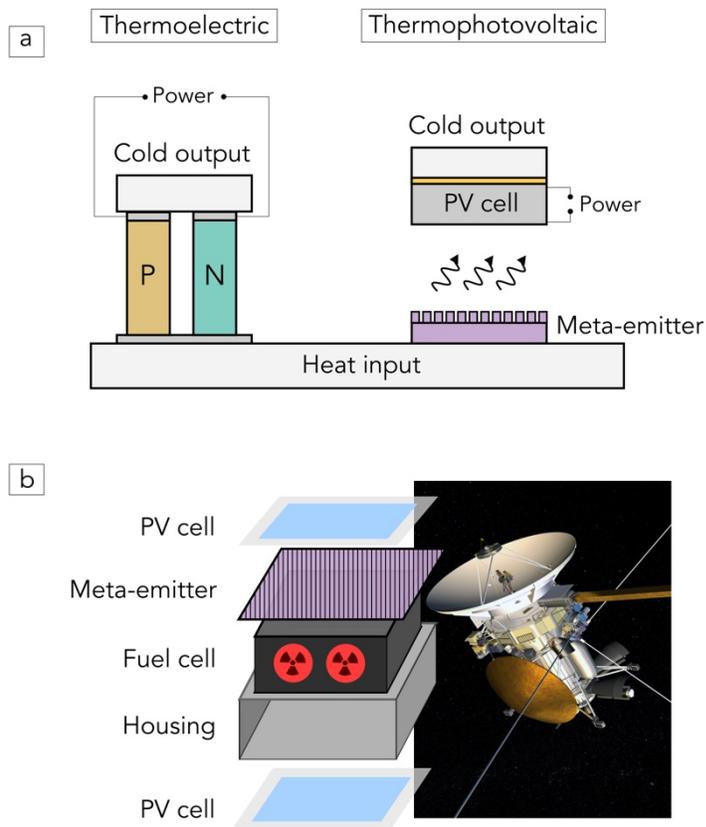

**Figure 3.** Converting heat into electricity. (a) Thermoelectric generators (left) operate based on the Seebeck effect, where a temperature gradient induces a voltage. On the other hand, in a thermophotovoltaic (TPV) converter (right), heat is radiated as light and converted into electricity by a photovoltaic cell. High-efficiency TPV conversion is possible when the spectrum of thermal emission is matched to the semiconductor absorption edge, with active ongoing research in metamaterial thermal emitters, high-quality photovoltaic cells, and high reflectivity mirrors. (b) Radioisotope-thermophotovoltaic (RTPV) device concept for high-efficiency heat-to-electricity conversion in space. (Credit: NASA)

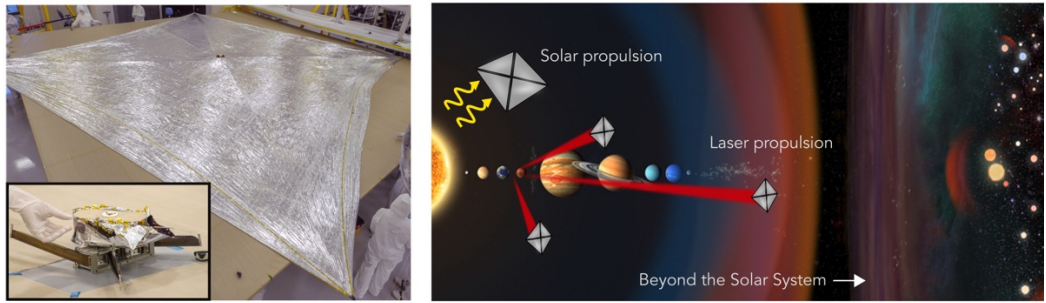

**Figure 4.** Photons for propulsion in space. (Left) Solar sail for the Near-Earth Asteroid (NEA) Scout mission under development by NASA. Compactly folded for transportation (inset), the sail will unfurl in space, covering an area of 86 m$^2$, and use sunlight to generate thrust by reflecting photons. (Right) Space exploration using light for propulsion. Smaller and much faster than solar sails, laser-driven spacecraft could be deployed in swarms, for space exploration within and beyond our solar system (e.g., Breakthrough Starshot Initiative[93]). Adapted with permission from NASA/JPL-Caltech.

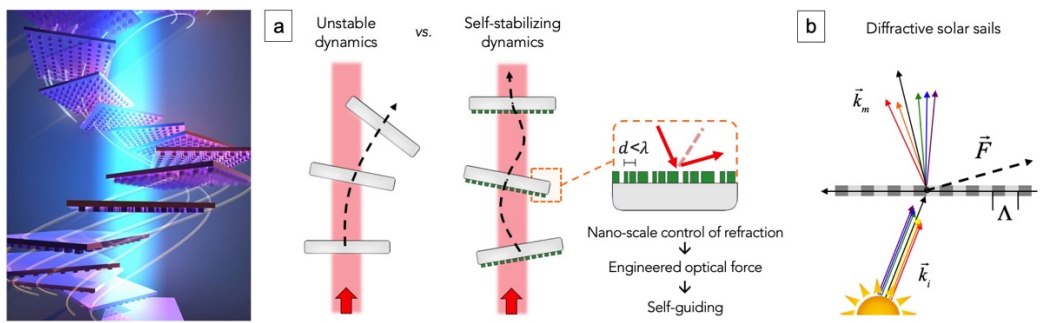

**Figure 5.** Optomechanical control with nanophotonic materials. (a) Conventional structures would be unstable in a light beam. In contrast, a lightweight structure embedded with subwavelength elements—a metasurface—could passively self-orient and self-stabilize. (b) Nanophotonic engineering of optical radiation forces could enable new mechanisms for trajectory control and propulsion of laser and solar-driven spacecraft. (Credit: G. Swartzlander)